\documentclass[12pt,english]{iopart}
\usepackage[T1]{fontenc}
\usepackage[latin1]{inputenc}
\usepackage{babel}
\usepackage{graphicx}

\begin{document}

\title{Cu-based metalorganic systems: an {\it ab initio} study of the electronic structure}

\author{L. Andrea Salguero$^1$, Harald O. Jeschke$^1$, Badiur Rahaman$^2$, Tanusri Saha-Dasgupta$^2$, Christian Buchsbaum$^3$, Martin U. Schmidt$^3$,  Roser Valentí$^1$ }

\address{$^1$ Institut für theoretische Physik, Johann Wolfgang
 Goethe-Universität, Max-von-Laue-Str. 1, 60438 Frankfurt/Main, Germany}

\address{$^2$ S.N. Bose National Centre for Basic Sciences, J.D Block,
Sector 3, Salt Lake City, Kolkata 700098, India}

\address{$^3$ Institut für Anorganische und Analytische Chemie, Johann Wolfgang
 Goethe-Universität, Max-von-Laue-Str. 7, 60438 Frankfurt/Main, Germany}

\ead{jeschke@itp.uni-frankfurt.de}


\begin{abstract}

Within a first principles framework, we study the electronic structure
of the recently synthesized polymeric coordination compound
Cu(II)-2,5-bis(pyrazol-1-yl)-1,4-dihydroxybenzene (CuCCP), which has
been suggested to be a good realization of a Heisenberg spin-1/2 chain
with antiferromagnetic coupling. By using a combination of classical
with {\it ab initio} quantum mechanical methods, we design on the
computer reliable modified structures of CuCCP aimed at studying
effects of \mbox{Cu-Cu} coupling strength variations on this spin-1/2
system.  For this purpose, we performed two types of modifications on
CuCCP. In one case, we replaced H in the linker by i) an electron
donating group (NH$_2$) and ii) an electron withdrawing group (CN),
while the other modification consisted in adding H$_2$O and NH$_3$
molecules in the structure which change the local coordination of the
Cu(II) ions.  With the NMTO-downfolding method we provide a
quantitative analysis of the modified electronic structure and the
nature of the \mbox{Cu-Cu} interaction paths in these new structures
and discuss its implications for the underlying microscopic model.


\end{abstract}




\submitto{\NJP}

\maketitle

\section{Introduction}\label{sec:introduction}

Low-dimensional quantum spin systems have been a subject of intensive research
in the last decades, especially in solid state theory.   
The discovery in recent years of low-dimensional materials has allowed
a direct comparison of experimental observations with theory
predictions. Out of these studies, the need of available
low-dimensional materials with a certain degree of tunability has
emerged since tunability has the advantage of allowing a step by step
understanding of the complexity of the materials. In this paper we
will focus on the search and study of tunable low-dimensional
materials.  We will restrict ourselves to a particular class of
systems, namely metalorganic compounds.

Metalorganic compounds formed by transition metal centres bridged with
organic ligands are being intensively discussed in the context of new
molecule-based magnets and electronic materials~\cite{kahn,
postnikov:04}. They constitute a class of tunable materials partly due to
their modular nature.  The modular setup has the advantage of allowing
the modification of relevant subunits chemically without changing the
subsequent crystal engineering.  Substitution of organic groups and
ligands in these systems play the role of doping in the search for
materials with desired magnetic interaction strengths and charge
carrier concentrations.

In the present work we pursue these ideas from a theoretical point of
view.  We consider a computationally feasible combination of classical
with quantum mechanical {\it ab initio} tools~\cite{jeschke:06} in
order to design and analyze new metalorganic compounds.  As an
example, we introduce systematic changes on existing metalorganic
materials in order to achieve desirable electronic or magnetic
properties in the modified new structures.  Such study i) allows for a
gradual understanding of the properties of these low-dimensional
systems ii) provides a guide to systematic synthesis in the
laboratory.

\begin{figure}

\includegraphics[width=\textwidth]{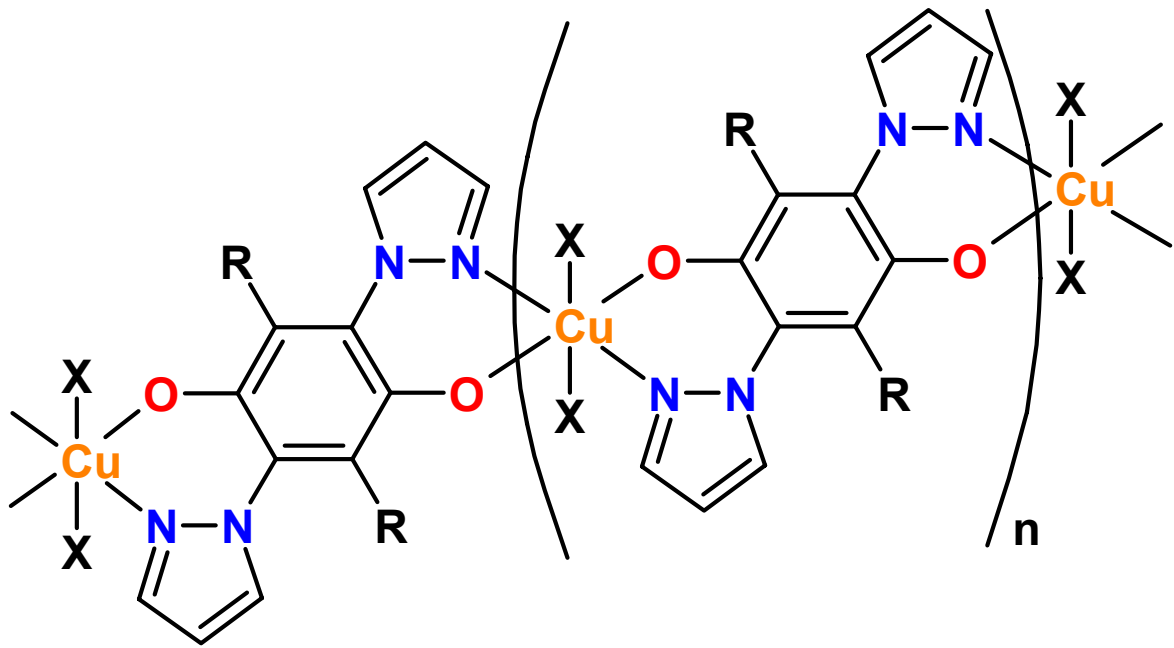}
\caption{Polymeric unit of Cu(II)-2,5-bis(pyrazol-1-yl)-1,4-dihydroxybenzene
  (CuCCP) (X=$\emptyset$, R=H). We will consider the substitutions R=CN and
  R=NH$_2$ on the central benzene ring and the ligands X=H$_2$O and X=NH$_3$.
}
\label{chemdiagr}
\end{figure}

We focus our attention on the recently synthesized coordination
polymer
Cu(II)-2,5-bis(pyrazol-1-yl)-1,4-dihydroxybenzene~\cite{dinnebier:02}
(CuCCP) which, from susceptibility measurements~\cite{wolf:04}, has
been identified as a model system for a spin-1/2 Heisenberg chain with
an antiferromagnetic exchange coupling constant of $J=21.5$~K. The
polymeric unit is shown in Fig.~\ref{chemdiagr}. This compound is a
good starting point to study effects of coupling strength variation by
appropriately introducing modifications on the linkers as well as
changes on the Cu coordination. Note that the coupling strength scale
for this system is small compared f.i. to inorganic Cu
oxides~\cite{Singh:02}, where the couplings are more than one order of
magnitude larger. Accordingly, in this work we will be expecting
coupling strength variations in the range of one to a few tens of meV
in energy.

The Cu{\textendash}Cu interaction in this compound depends on the electronic
nature of the linker. Its properties can be tuned smoothly and
predictably by changing the substitution pattern R (see Fig.\
\ref{chemdiagr}) of the central benzene ring (hydroquinone), or by
introduction of additional ligands X at the Cu(II) ions. The substitution
or introduction of additional ligands is expected to bring changes
in the electronic properties of the compound. For example, it can
change the magnitude of magnetic interactions between the Cu(II) 
centres in the spin chain via  change in the charge density in the 
polymeric chain. It may change the effective, inter-chain interactions,
the one-dimensional nature of the original compound may thus be modified.
It may even change the coordination and valence of the Cu(II) ions, which
may induce changes in the transport properties along the
one-dimensional chain by moving away from the Mott insulator at a
half-filled Cu~$3d_{x^2-y^2}$ orbital.

With the above ideas in mind, we first considered two possible H
replacements in the central benzene ring: an NH$_2$ group, which acts
as electron donating group, and a CN group, which acts as electron
withdrawing group (see Fig.\ \ref{chemdiagr}). Furthermore, extensive
crystallization trials showed that crystallites of the Cu(II) polymer
always contain lattice defects in high concentration. In many
compounds, the Cu(II) ions are coordinated by six nitrogen or oxygen
atoms instead of four ligands. The Cu(II) polymer is experimentally
crystallized in a mixture of water and ammonia solvents, and it is
likely that H$_2$O or NH$_3$ molecules are built into the crystal
lattice. We have, therefore also considered the introduction of
additional ligands like H$_2$O or NH$_3$ at the Cu(II) ions in our
simulation study.  In Fig.\ \ref{chemdiagr} and Table
\ref{structure_table} we give an account of these modified structures.

The paper is organized as follows: In section \ref{sec:method} we give
an account of the computational methods considered in this work.  In
section \ref{sec:results} we present our results and discussion of the
relaxed new crystal structures and the analysis of their electronic
properties. Finally in section \ref{sec:summary} we summarize our
findings.

\begin{table}
\renewcommand{\arraystretch}{1.1}
\renewcommand{\tabcolsep}{10pt}
\begin{center}
\begin{tabular}{l|c|c}
Short name&R&X\\\hline
CuCCP&H&-\\
Cu(II)-NH$_2$&NH$_2$&-\\
Cu(II)-CN&CN&-\\
Cu(II)-H$_2$O&H&H$_2$O\\
Cu(II)-NH$_3$&H&NH$_3$
\end{tabular}
\end{center}
\caption{Naming convention for the substitutions and ligands on the CuCCP
  coordination polymer considered in this work.}
\label{structure_table}
\end{table}

\section{Method}\label{sec:method}

The methods used in this work, can be primarily categorized into two
classes. Firstly, a class of methods has been used for the accurate
structural determination of both the parent and the modified
compounds. Once the structural aspects are decided, their electronic
structures are calculated and analyzed with another class of
methods. Note that the understanding of a complex system and design
of new compounds need a combination of several different methods, each
being focused to deal with one specific aspect. In the following, we
give a brief description of all the methods that we have employed.

\subsection{Determination of Crystal Structure}\label{sec:method_structure}

In the absence of diffraction data, a method much used to {\it a
priori} predict crystal structures, is the force field
technique~\cite{schmidt:96,verwer:98,mooij:98,lommerse:00,motherwell:02,day:05}.
While such calculations are computationally fast, they rely on a
classical ansatz and therefore miss all possible quantum mechanical
effects, which are important for the description of the electronic
structure.  Quantum mechanical methods, on the other side, are
computationally much more demanding, and they are, in this context,
typically employed for two tasks: One is the local optimization after
global optimization with force field methods. This has been reported
for inorganic systems like NaCl or MgF$_2$~\cite{schoen:01a,schoen:01b} and for
simple organic compounds like glycole C$_2$H$_4$(OH)$_2$ and glycerol
C$_3$H$_5$(OH)$_3$~\cite{eijck:01}. Another is for secondary
computations like the determination of molecular geometries,
electrostatic charges or for the calculation of intramolecular and
intermolecular potential curves~\cite{mooij:99a,mooij:99b}.

In the present work we used an effective way of designing reliable
crystal structures which shares the advantages of both methods, namely
the fast calculations with classical force field methods and the
subsequent accurate quantum mechanical description with {\it ab
initio} methods. We first created the modified structures on the basis
of crystallographic databases~\cite{cambridge} and the crystal
structures were optimized by force field methods. In the second step
the structures were relaxed by {\it ab initio} quantum mechanical
molecular dynamics~\cite{car:85} within the density functional theory
formalism (DFT) until the forces on the atoms were less than a given
threshold to ensure structure stability. Our work differs from the
mentioned previous works in the sense that using this approach, we
succeeded in treating materials with large unit cells (of the order of
100 atoms) and complicated electronic structure (transition metal
complexes) with sufficient accuracy.

All force field optimizations were performed using the program package
Cerius2~\cite{cerius}. We modified the Dreiding 2.21~\cite{mayo:90}
force field by introducing energy terms for the case of octahedrally
coordinated metal ions. For the energy minimizations we used the modified
Dreiding force field with Gasteiger~\cite{gasteiger:80} charges. All
structural models were based on the experimentally determined crystal
structure of CuCCP~\cite{dinnebier:02}. The crystallographic
symmetry of the structure models was maintained in all
relaxations. The position of the Cu ion was kept fixed 
during all force field and quantum mechanical optimizations.
The second step of quantum mechanical relaxations were performed 
by Car Parrinello (CP) {\it ab initio} molecular dynamics (AIMD) 
calculations~\cite{car:85} based on the Projector Augmented Wave (PAW) 
method~\cite{bloechl:94,note}.

\subsection{Electronic Structure Calculations}\label{sec:method_elec}

We computed the electronic structure of the relaxed structures with
 the
Full Potential Linearized Augmented Plane Wave basis (FPLAPW) as
implemented in the Wien2k code~\cite{wien}. Calculations were
performed within the Generalized Gradient Approximation (GGA)~\cite{perdew:96}.
The choice of muffin-tin radii $r_\mathrm{MT}$,
$k$ mesh and plane-wave cutoffs $k_\mathrm{max}$ were carefully
tested. We considered a $k$ mesh of $(8 × 6 × 5)$ in the irreducible
Brillouin zone and a $Rk_\mathrm{max}=3.8$, which is reasonable for
systems that contain hydrogen atoms.

It  is well known that LDA or GGA fails to 
describe the correct insulating ground state for strongly correlated electron
system, as is the case here. Introduction of missing correlation
effects in a static mean-field like treatment as is done in the so called
LDA+U approach\cite{andersen:91,petukhov:03}, should give
rise to the correct insulating state, as is supported by our calculations
(not shown here). In the
present paper we are mainly interested in estimating the effective
one-electron hopping interactions which are well
described within LDA or GGA. In fact, the use of DFT calculations
to understand the chemistry of correlated materials
is a well established method~\cite{pavarini:05}.

Finally, in order to analyze the computed electronic structure and to
extract an effective microscopic Hamiltonian, we derived
quantitatively the \mbox{Cu{\textendash}Cu} hopping integrals within the
NMTO-{\it downfolding} technique~\cite{nmto1,nmto2} as implemented in the Stuttgart
code~\cite{stuttgart,convergence}.  The {\it downfolding} technique provides
a useful way to derive few-orbital Hamiltonians starting from a
complicated full LDA Hamiltonian by integrating out degrees of freedom
which are not of interest. This procedure naturally takes into account
the renormalization effect due to the integrated-out orbitals by
defining energy-selected, effective orbitals which serve as
Wannier-like orbitals for the few-orbital Hamiltonian in the {\it
downfolded} representation. The method provides a first-principles way
for deriving a few-band, tight-binding Hamiltonian of the form $H_{TB}
= \sum_{ij} t_{ij}(c^{\dagger}_i c_j + h.c.)$ for a complex system, where the
$t_{ij}$ define the effective hopping between the downfolded orbitals
and $c^{\dagger}_i$ ($c_i$) are electron creation (annihilation) operators
on site $i$. This method has proved to be extremely successful for
systems such as high-Tc cuprates~\cite{pavarini:01}, double
perovskites~\cite{sarma:00} or low-dimensional inorganic quantum spin
systems~\cite{valenti:01,valenti:02a,valenti:02b,sahadasgupta:02,valenti:03}.

Such estimates of the effective hopping integrals are useful in
defining the underlying low-energy magnetic model. More precisely, the
one-electron effective Cu-Cu hopping integral, $t$ can be related to
the Cu-Cu magnetic exchange coupling interaction $J$ via a
second-order perturbative treatment within the framework a many-body
Hubbard-like model.  Assuming that these couplings are
antiferromagnetic and neglecting ferromagnetic contributions, $J$ can
be estimated as $J_\mathrm{AFM}\approx4t^2/U_\mathrm{eff}$ where $U_\mathrm{eff}$
is the effective onsite Coulomb repulsion on the Cu site.

\section{Results}\label{sec:results}

\subsection{Crystal Structure}\label{sec:structure}

\begin{figure}
\begin{center}
\includegraphics[width=0.55\textwidth,bb=250 0 1315 1132,clip]{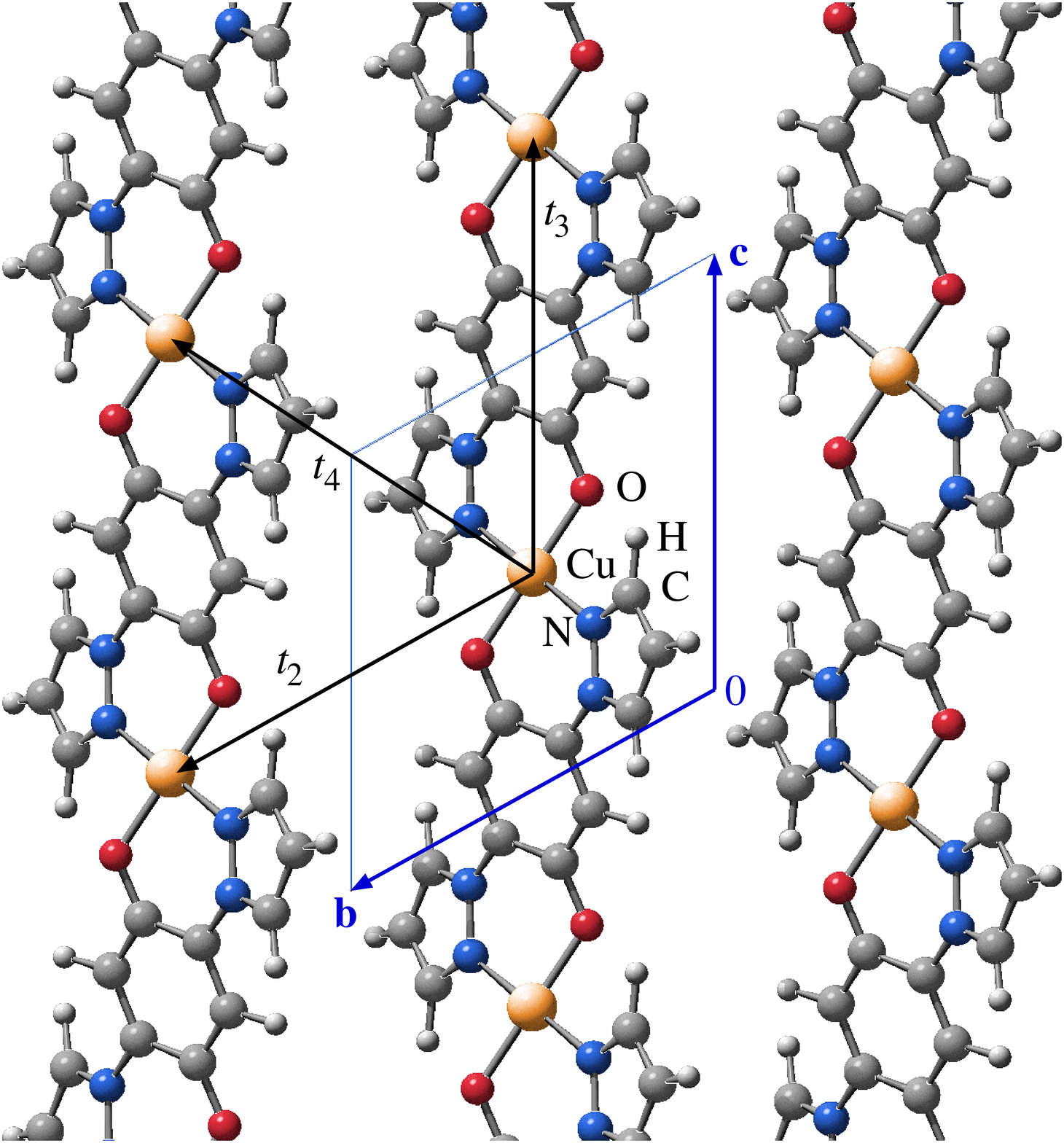}
\vspace{0.7cm}

\includegraphics[width=0.45\textwidth]{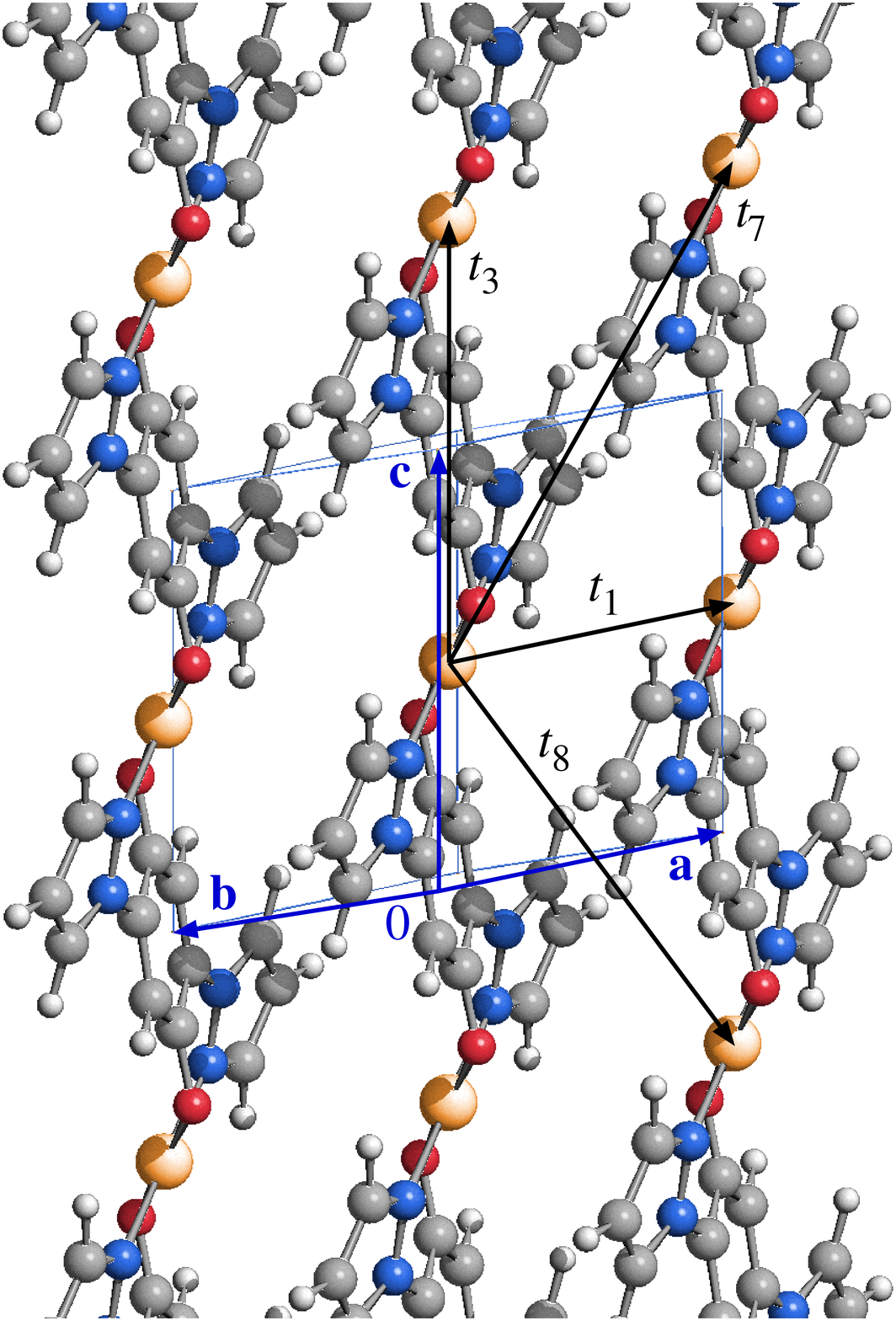}
\end{center}
\caption{ 
Crystal structure of the CuCCP polymer in two different
orientations. The unit cell is shown in the figures (vectors a, $b$,
and $c$).  Note the arrangements of the Cu chains along the $c$
direction. The various \mbox{Cu{\textendash}Cu} interaction paths $t_i$ have
been also drawn where the index $i=1,2,3,7,8$ denotes the $i$-th
nearest neighbour.}
\label{path1}
\end{figure}

Dinnebier {\it et al.}~\cite{dinnebier:02} reported the synthesis, and
crystal structure determination of the CuCCP system obtained by
layering a solution of 2,5-bis(pyrazol-1-yl)-1,4-dihydroxybenzene in
CH$_2$Cl$_2$ with a solution of CuBr$_2$ in concentrated aqueous
ammonia. The system crystallizes in the triclinic space group
$P\bar{1}$ with 27 atoms per unit cell. This compound tends to form
independent polymeric chains consisting of deprotonated
2,5-bis(pyrazol-1-yl)-1,4-dihydroxybenzene molecules bridged by Cu(II)
ions with a 3$d^9$ configuration, which corresponds to a local spin
1/2. As shown in Fig.\ \ref{path1} the chain axes are oriented along
the $c$ axis of the crystal and the copper ions are located at $(1/2,
1/2, 1/2)$, which is a centre of symmetry of the space group
$P\bar{1}$. The Cu{\textendash}Cu distance along the approximate $a$ axis is
about 5.2~Å while along the other two axes it is close to 8~Å.  These
Cu(II) ions are coordinated in an almost square planar fashion by two
(pyrazolyl) nitrogen atoms and two oxygen atoms of the deprotonated
dihydroxybenzene groups.
   
The crystal structure was determined from X-ray powder diffraction
data; consequently the overall structure and the arrangement of the
chains are reliable, but the individual atomic positions had an
accuracy of only about 0.3~Å.  A DFT analysis of the
forces~\cite{wien} between the atoms shows that the experimentally
determined CuCCP structure is still very unstable with forces of the
order of 100~mRyd/$a_\mathrm{B}$ or more for some atoms. We have
therefore relaxed the atomic positions keeping the Cu position fixed
with the AIMD method described in section \ref{sec:method_structure}.

The Cu(II)-NH$_2$ polymer and the Cu(II)-CN polymer were generated
from the CuCCP polymer, by substituting the two hydrogen atoms of the
benzene rings by amino (NH$_2$) or cyano (CN) groups, respectively
(see Fig.~\ref{chemdiagr} and Table \ref{structure_table}). The
Cu(II)-H$_2$O (Cu(II)-NH$_3$) polymer was constructed from the CuCCP
polymer by adding two water molecules (ammonia molecules) as
additional ligands to the Cu(II) ion (see Figs.~\ref{chemdiagr} and
and Table \ref{structure_table} ). In the original crystal structure
the chains are quite densely stacked. The introduction of the H$_2$O
(or NH$_3$) molecules would either lead to unrealistically short
contacts to the neighbouring chains, or to a considerable increase of
the distances between the chains, resulting in an unrealistically
loosely packed structure. Therefore the crystal structures of the
Cu(II)-H$_2$O and Cu(II)-NH$_3$ polymers were fully optimized,
including an optimization of the lattice parameters.  Moreover, in
order to achieve a better packing of the Cu(II) polymer chains with a
favourable lattice energy, the Cu(II) chains shifted in the
optimization process both sidewards as well as along the chain
direction with respect to each other. The resulting unit cell
parameters are shown in Table~\ref{latpam}.

\begin{table}
\caption{Lattice parameters for the experimental crystal structure CuCCP
and models Cu(II)-H$_2$O and Cu(II)-NH$_3$, optimized with force field
methods.} 

\begin{center}
\begin{tabular}{c c c c c c c c} \hline
Structure &$a ($Å$)$&$b ($Å$)$&$c ($Å$)$&$\alpha ($°$)$&$\beta ($°$)$&$\gamma ($°$)$&$V($Å$^3)$\\ \hline 
{CuCCP} exp & $5.172$ & $7.959$ & $8.230$ & $118.221$ & $91.520$ & $100.148$ & $291.47$ \\
{Cu(II)}-H$_2$O & $5.234$ & $11.249$ & $8.072$ & $117.611$ & $68.822$ & $127.155$ & $330.43$ \\
{Cu(II)}-NH$_3$ & $5.459$ & $11.597$ & $8.349$ & $118.423$ &
$68.840$ & $130.883$ & $350.49$ \\ \hline
\end{tabular}
\end{center}
\label{latpam}
\end{table}

All the modified structures were relaxed in the second step with the
AIMD method until the forces on the atoms were sufficiently small to
ensure stability of the quantum mechanical calculations. In the
Appendix we present the relaxed structural data of all the modified
structures.

\subsection{Electronic Structure and Effective Cu-Cu Interactions}\label{sec:electronic}

\begin{figure}
\begin{center}
\includegraphics[width=0.8\textwidth]{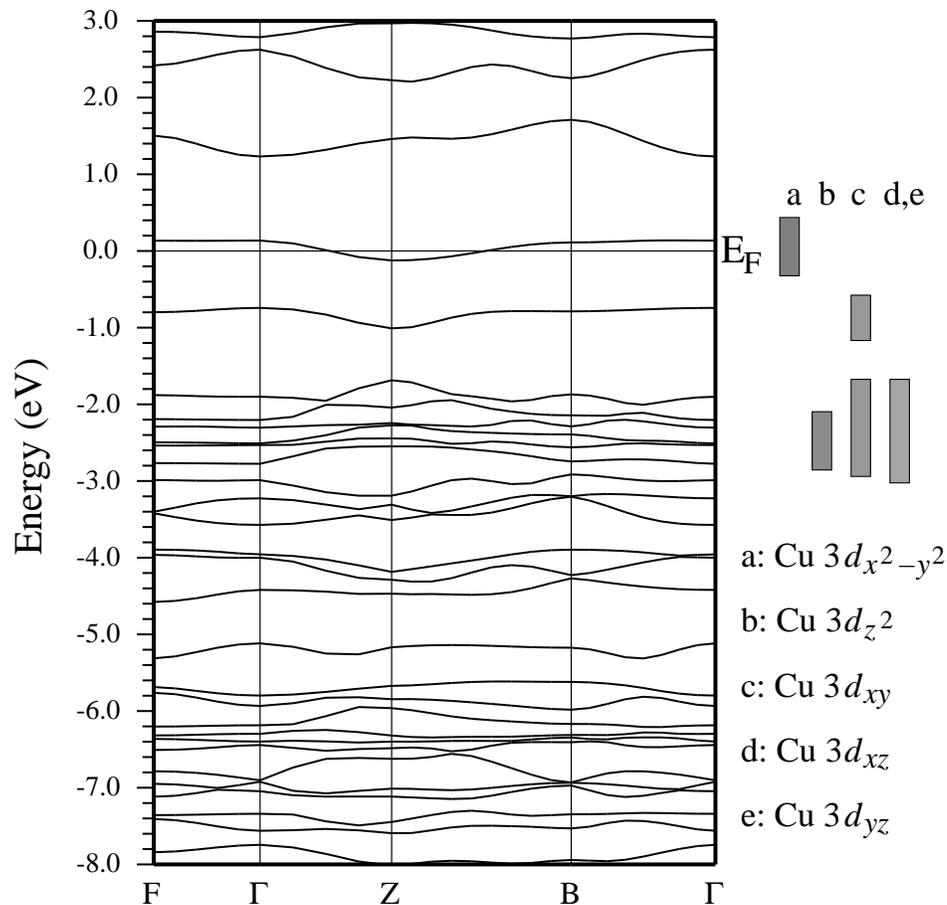}
\end{center}
\caption{
Band structure for the relaxed Cu(II) polymer CuCCP in the GGA
approximation along the path~\cite{bradley72}
F$(0,1,0)$-$\Gamma(0,0,0)$-Z$(0,0,1)$-B$(0.99, -0.13, 0)$-$\Gamma(0,0,0)$ in units of $\pi/a$, $\pi/b$,
$\pi/c$. The bars indicate the dominant band character in the local coordinate
frame of Cu (see text for explanation).}
\label{CuII-orig-spag-rang2-2} 
\end{figure}

\subsubsection{CuCCP}

In Fig.~\ref{CuII-orig-spag-rang2-2} we present the band structure for
the relaxed CuCCP where the Cu~$d$ band character is shown by bars on
the right side. The band characters are given in the local coordinate
frame of Cu which is defined with the local $z$~direction pointing
from the Cu to out-of-plane N atom in the next layer and the
$y$~direction pointing from the Cu to in-plane O atom. Cu is in a
3$d^9$ configuration, with all $d$ bands occupied except for the last
band which is half-filled.  GGA predicts a metallic behaviour for this
system. As mentioned previously, inclusion of on-site electronic
correlation within LDA+U opens a gap between a lower occupied Hubbard
band and an upper unoccupied Hubbard band and the system is described
as a Mott-Hubbard insulator. Since the O-Cu-N angle in the
CuO$_2$N$_2$ plane is not exactly 90°, the various Cu~$d$ degrees of
freedom defined with respect to the local coordinate frame mentioned
above show slight admixtures. In particular, the Cu~$d_{x²-y²}$
dominated band crossing the Fermi level has also small contributions
from Cu~$d_{yz}$ degrees of freedom which arise from the distorted
geometry.

>From the dispersion of the Cu~$d$ band at the Fermi level we
confirm the one-dimensionality of the structure. The paths F-$\Gamma$ and
B-$\Gamma$ which correspond to the inter-chain paths are almost
dispersionless and the intrachain $\Gamma$-Z-B path shows a cosine-like
behaviour. A quantitative analysis of the various hopping integrals
obtained with the downfolding procedure, by keeping only the
Cu~$d_{x²-y²}$ degrees of freedom active and integrating out all the
rest within the NMTO framework, is given in the first column of
Table~\ref{hopping}. The various interaction paths are as shown in
Fig.~\ref{path1}. The largest hopping integral $t_3$ is along the
chain (see Fig.~\ref{path1}) while all other hoppings are almost an
order of magnitude smaller.

\begin{table}
\caption{ Values for the \mbox{Cu{\textendash}Cu} hopping integrals
calculated with the NMTO downfolding method for the relaxed CuCCP,
Cu(II)-NH$_2$ and Cu(II)-CN structures. The values are given in
meV. The subscripts $i$= 1,2,3,7,8,12 denote the $i$th nearest
neighbours. See Fig. \protect\ref{path1}. Only the hopping integrals
having values larger or equal to one tenth of a meV has been shown.}

\begin{center}
\begin{tabular}{|c|c|c|c|}\hline
Path & CuCCP & Cu(II)-NH$_2$ & Cu(II)-CN \\ \hline \hline
$t_1$ & 4 & 9 & 22 \\
$t_2$ & 8 & 3 & 0\\
$t_3$ & 79 & 88 & 68\\
$t_7$ & 5& 1 & 9 \\
$t_8$ & 3 & 8 & 8\\
$t_{12}$ &0 &0 &9\\ \hline
\end{tabular}
\end{center}
\label{hopping}
\end{table}

\subsubsection{Cu(II)-NH$_2$ and Cu(II)-CN}

In order to have a quantitative account of the structural changes that
the polymer system undergoes under the various substitutions, we
define the angle between the vector perpendicular to the CuO$_2$N$_2$
plane and the vector perpendicular to the benzene ring as the tilting
angle $\vartheta$. The substitution of H by NH$_2$ groups or CN groups
in the benzene rings induces a tilting from $\vartheta=34.9$° in CuCCP
to $\vartheta=37.3$° in Cu(II)-NH$_2$ and $\vartheta=36.3$° in
Cu(II)-CN.

\begin{figure}
\begin{center}
\includegraphics[angle=-90,width=0.74\textwidth]{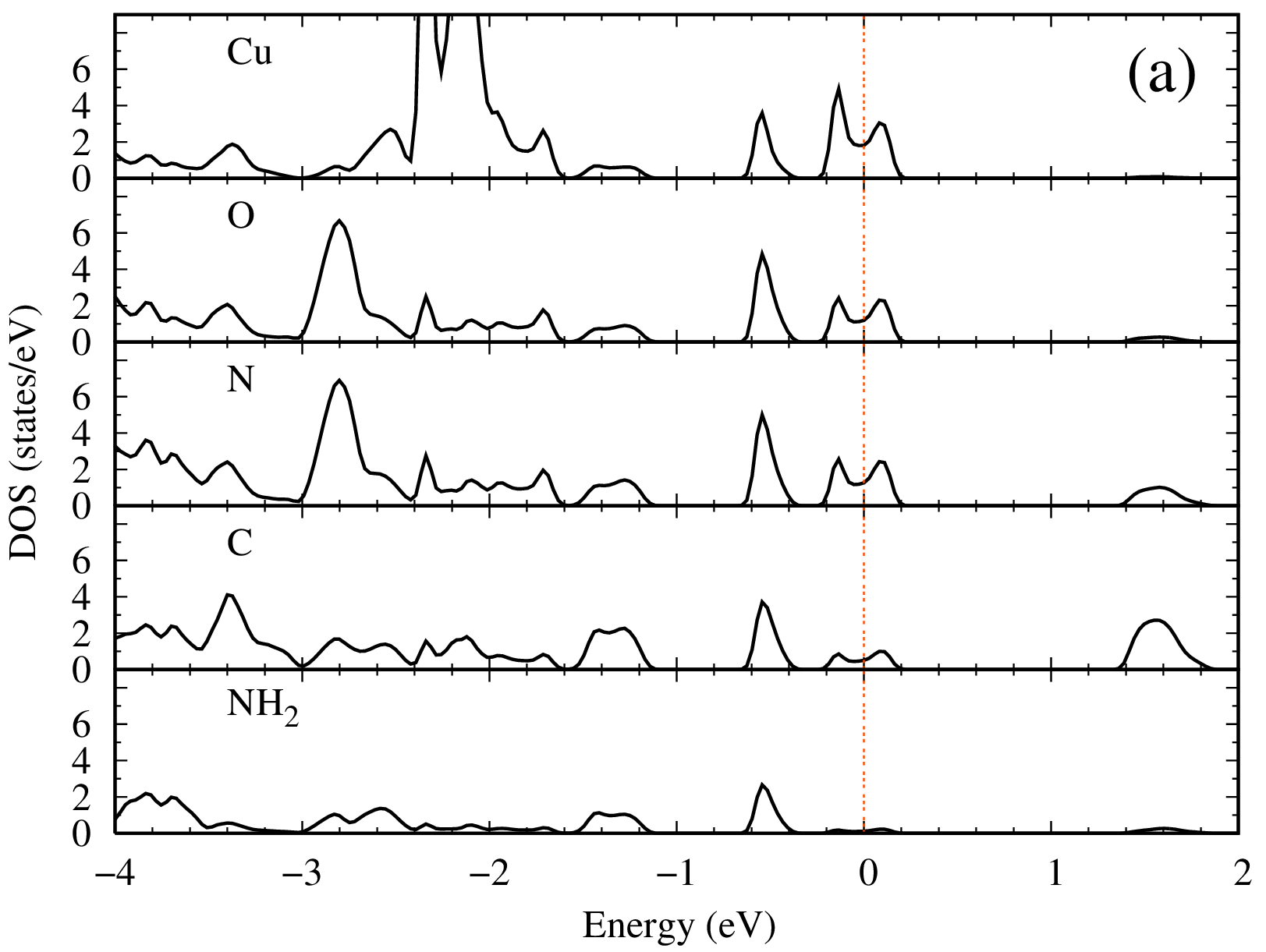}
\includegraphics[angle=-90,width=0.74\textwidth]{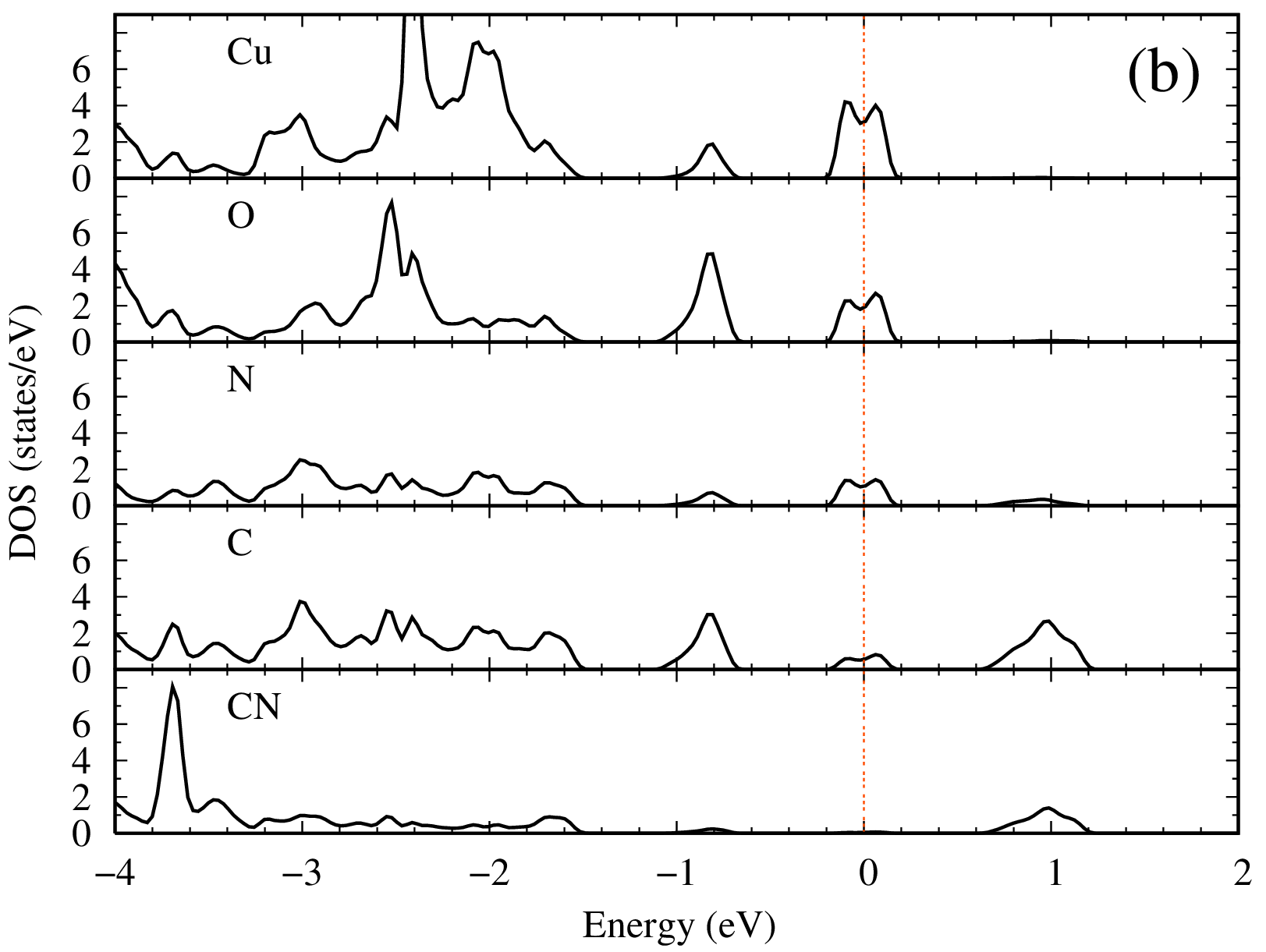}
\includegraphics[angle=-90,width=0.73\textwidth]{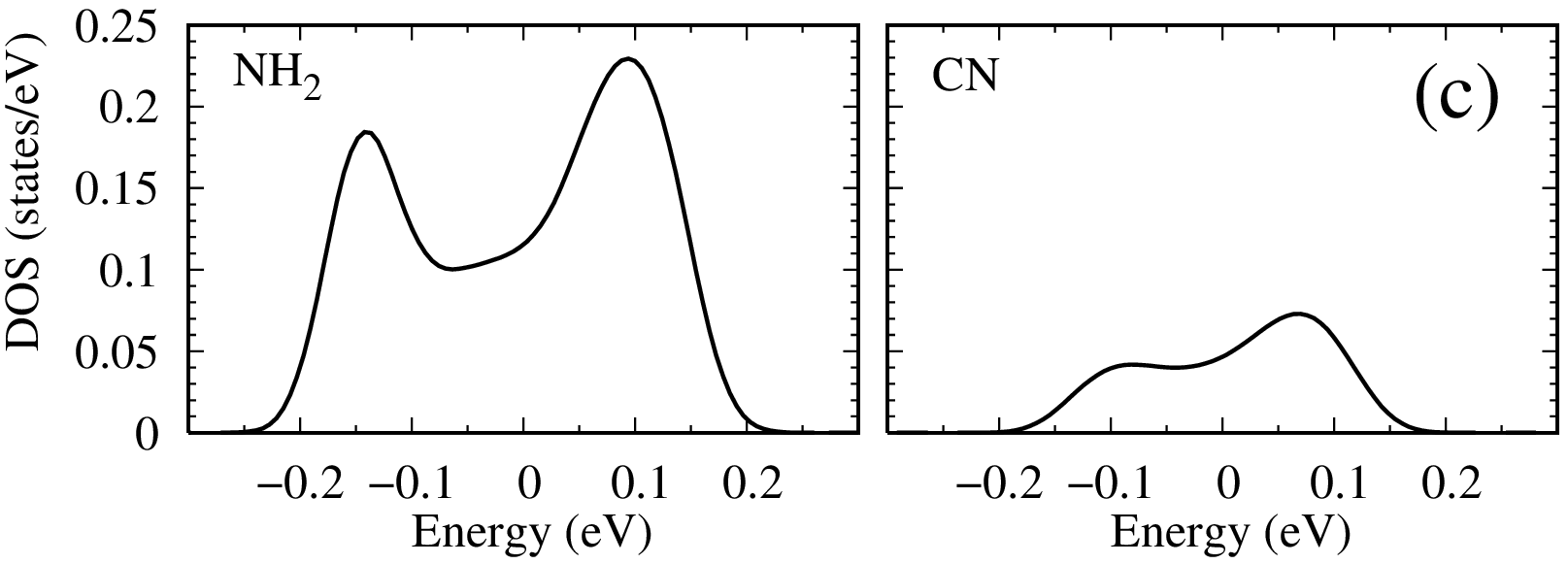}
\end{center}
\caption{
Orbital resolved DOS for (a) Cu(II)-NH$_2$ and (b) Cu(II)-CN. 
Panel (c) shows the comparison of the contribution of the NH$_2$
and CN groups to the DOS at $E_\mathrm{F}$ in a blown up scale.  }
\label{dos-comparison} 
\end{figure}

In Figs.~\ref{dos-comparison} (a) and (b) we present the FPLAPW
orbital resolved DOS for the Cu(II)-NH$_2$ and the Cu(II)-CN within
the GGA approximation. Shown is the contribution to the total DOS of
Cu, O, N, C and the groups NH$_2$ and CN. While some changes in the
detail shape of the DOS for Cu, O, N and C between
Figs.~\ref{dos-comparison} (a) and \ref{dos-comparison} (b) are
observed, the most important effect is the different electronic nature
of the NH$_2$ and CN groups. Fig.~\ref{dos-comparison} (c) shows the
contribution of these groups at the Fermi surface.  The CN group bands
are deep down into the valence band while the NH$_2$ group has
appreciable contribution near the Fermi level, which indicates its
involvement in the effective interaction paths between copper
atoms. We will see this more clearly in the plot of the NMTO-Wannier
orbitals to be discussed later in this section.

\begin{figure}
\begin{center}
\includegraphics[height=0.7\textwidth,angle=-90]{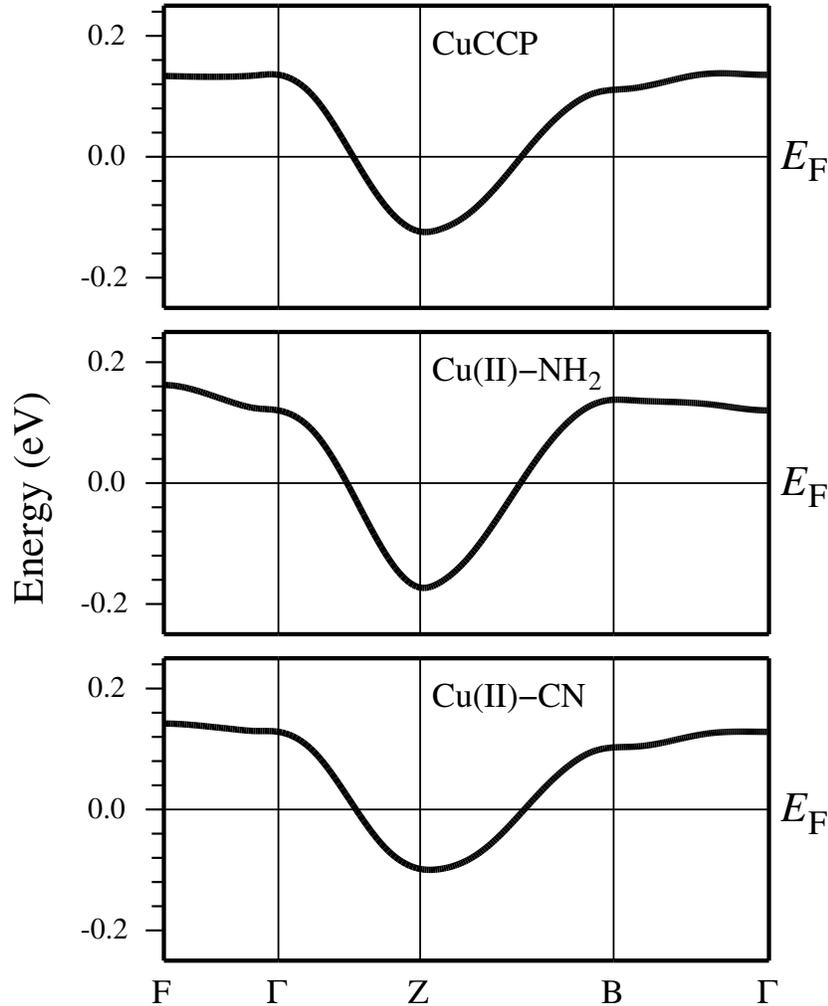}
\end{center}
\caption{ 
Comparison between the band structures for (from top to bottom) the
relaxed CuCCP, Cu(II)-NH$_2$ and Cu(II)-CN respectively.}
\label{CuII-orig-subs}
\end{figure}

In Fig.~\ref{CuII-orig-subs} we show a comparison of the band
structure for the relaxed CuCCP, Cu(II)-NH$_2$ and the Cu(II)-CN
polymers in the energy range $[-0.25~\mathrm{eV}, 0.25~\mathrm{eV}]$ where
only the Cu~$d_{x^2-y^2}$ dominated band is involved. Though the basic
nature of the dispersion remains the same upon substitution, the
details however do change. NH$_2$ seems to be the most effective
substitution to increase the intrachain \mbox{Cu{\textendash}Cu} interaction
(the bandwidth widens along the $\Gamma$-Z-B path for the Cu(II)-NH$_2$
system). The CN substitution, on the other hand, reduces this
interaction (note the bandwidth narrowing along the $\Gamma$-Z-B path for
the Cu(II)-CN system). The substitution process also enhances certain
interchain couplings. The almost dispersionless behaviour along F-$\Gamma$
and B-$\Gamma$ becomes more dispersive. Description of such fine and subtle
changes, need some quantitative measures which can be best described
by the changes in effective Cu-Cu hoppings.  This is shown in
Table~\ref{hopping} where the hopping integrals obtained by the NMTO
downfolding method are shown. Note that the $t_1$ hopping for the
Cu(II)-CN system along the crystallographic $a$ direction is enhanced
by a factor of 4.5. Similarly, $t_7$ and $t_8$ hopping terms for the
Cu(II)-CN system between neighbouring Cu chains in the $b$ direction
(see Fig.~\ref{path1}) are almost 2-3 times larger compared to that of
the CuCCP system. The long-ranged $t_{12}$ hopping parameter between
neighbouring chains along the $a$ axis also attains appreciable
enhancement compared to a vanishing small value for the CuCCP system.
Similarly, $t_1$ and $t_8$ hoppings are enhanced for the NH$_{2}$
substitution by factors of about $\approx$ 2-3. Among all the hoppings, only
$t_2$ shows the exception of being systematically decreased upon
substitution. The predominant hopping, $t_3$ is enhanced in the
Cu(II)-NH$_2$ system and reduced in the Cu(II)-CN system as already
predicted from bandwidth arguments.

\begin{figure}
\begin{center}
\includegraphics[width=0.75\textwidth]{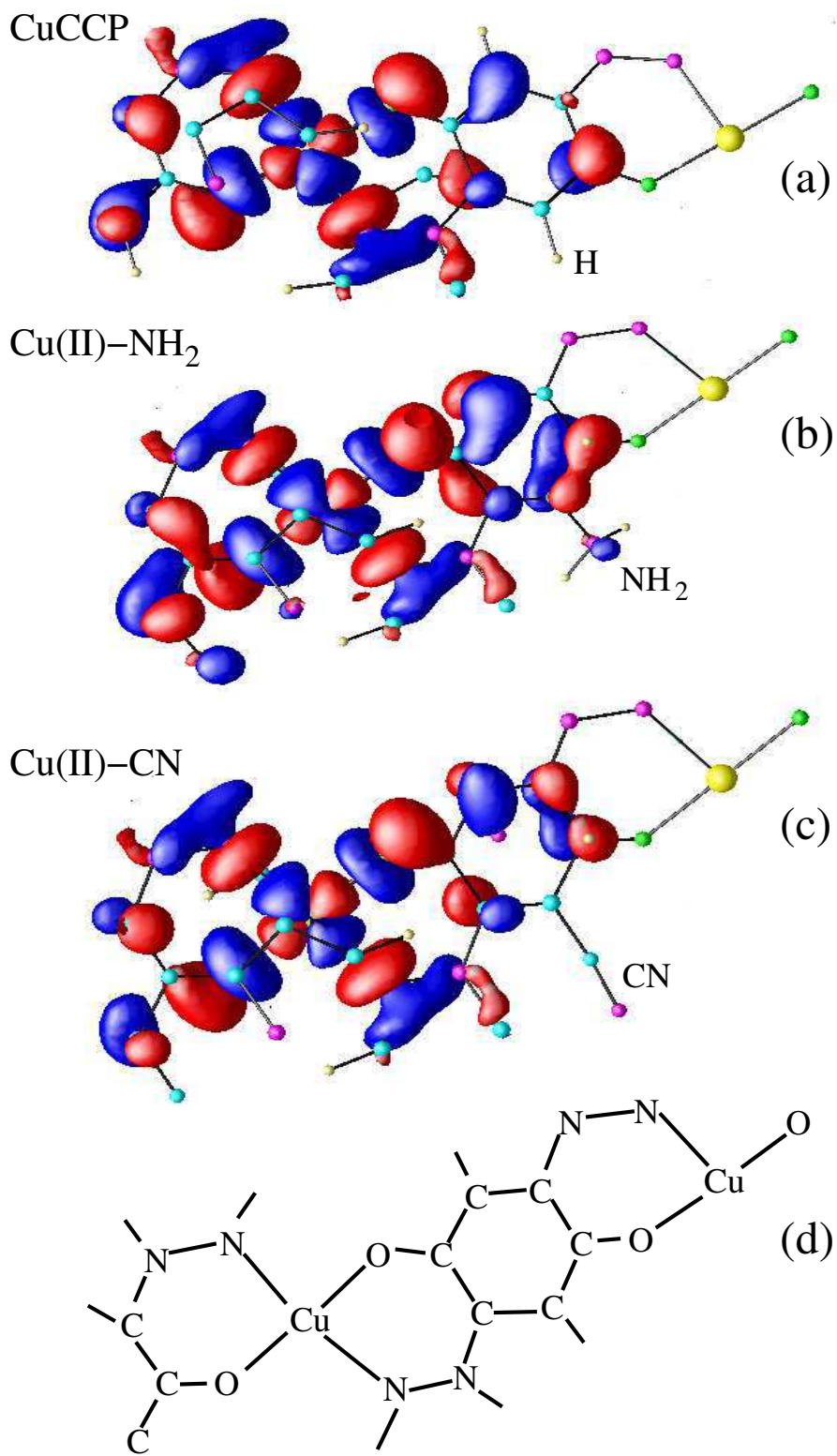}
\end{center}
\caption{ 
Cu Wannier functions for (a) the relaxed CuCCP polymer, (b) the
Cu(II)-NH$_2$ polymer, and (c) the Cu(II)-CN polymer. (d) indicates
the atom positions common to (a)-(c). The N-C-C-C-H chain of atoms
appearing above the Wannier function belongs to the next layer.}
\label{wannier}
\end{figure}

A very helpful tool to understand the origin of these changes is the
plot of effective Cu Wannier orbitals. In Fig.
\ref{wannier} we show the Wannier orbitals for the three Cu(II) systems
presented so far. The plotted Wannier orbitals are obtained with the NMTO downfolding
technique as introduced in the section \ref{sec:method_elec}.


The effective Cu Wannier orbital has the expected Cu~$d_{x^2-y^2}$
symmetry at the central Cu site while the tails sitting at other sites
are shaped according to the symmetries of the various integrated out
orbitals like the rest of Cu~$d$ orbitals, O~$p$, N~$p$ or C~$p$.
Note that the Cu~$d$, O~$p$ and N~$p$ antibonding orbitals in the
basic CuN$_2$O$_2$ square plaquette remain similar in all three cases
but the effective orbital distribution in the benzene ring is markedly
different.
The changes are most prominent for the NH$_2$ substituted
case with the tails attaining appreciable weight at the sites in the
benzene ring. We also notice the occurrence of weight at the NH$_2$
assembly which is in accordance with the orbital resolved DOS study
(see Fig.\ \ref{dos-comparison}). This leads to an enhancement in both
intra- and some interchain
\mbox{Cu{\textendash}Cu} interactions, caused by the larger overlap of the effective
orbitals. The enhancement happens via two different routes: one is due
to the different tilting of the benzene ring compared to the original
compound and the other one is the opening of additional interaction
paths via the NH$_2$ group which enhances the intrachain as well as
interchain interactions $t_1$ and $t_8$ as can be seen in the
quantitative estimates of the hopping interactions in
Table~\ref{hopping}. In the case of the CN substitution, the opening
of an additional intrachain pathway is absent, which is reflected  in the
reduced  intrachain ($t_3$)  hopping interaction. However the mechanism via the
tilting of the benzene ring is still operative which is seen  in
the enhancement of several interchain coupling, specially $t_1$.

Relating the magnetic coupling interaction $J$ with the effective
hopping interaction $t$ via a relationship
$J_\mathrm{AFM}\approx4t^2/U_\mathrm{eff}$, as discussed in
Section~\ref{sec:method_elec}, and choosing $U_\mathrm{eff}$ to be 5 eV
\cite{comment_U} we obtain the nearest neighbour coupling for CuCCP
system to be $J_\mathrm{AFM} \approx58$~K which is somewhat larger than the
experimental estimate~\cite{wolf:04} obtained by fitting
susceptibility data to an effective nearest neighbour Heisenberg model,
but remains of the same order of magnitude. The $J_\mathrm{AFM}$ values
estimated for Cu(II)-NH$_2$ and Cu(II)-CN systems are $J_\mathrm{AFM} \approx
72$~K and $\approx 43$~K respectively.

\subsubsection{Cu(II)-H$_2$O and  Cu(II)-NH$_3$}

 In our second set of modifications we introduce two kinds of ligands,
 H$_2$O and NH$_3$ in the CuCCP system in the way presented in
 Section~\ref{sec:structure}. Our goal is to study the effect
 of H$_2$O and NH$_3$ satellites on the CuCCP structure as well as to
 search for possible routes to change the Cu coordination from planar
 to octahedral. As explained in Section~\ref{sec:structure}, in
 order to obtain realistic structures, the optimization with the force
 field method was done without keeping the original cell fixed, since
 that would force very short intermolecular distances between the
 H$_2$O (NH$_3$) moieties and the neighbouring chains. The force field
 optimized structures (see Table~\ref{latpam}) and subsequently
 relaxed with AIMD are characterized by Cu-O (O of the H$_2$O
 molecule) distances of $d_\mathrm{CuO}=2.17$~Å while the Cu-O and Cu-N
 in-plane distances are $d_\mathrm{CuO}=1.99$~Å and
 $d_\mathrm{CuN}=2.01$~Å, respectively. This corresponds to a distorted
 octahedron elongated along the Cu-H$_2$O direction. For the case of
 the NH$_3$ ligands the Cu-N (N of the NH$_3$ molecule) distances are
 $d_\mathrm{CuN}=2.14$~Å, while the Cu-O and Cu-N in-plane distances are
 $d_\mathrm{CuO}=2.02$~Å and $d_\mathrm{CuN}=2.03$~Å, also giving rise to
 an elongated octahedron along the Cu-NH$_3$ direction. The ligands
 close to the Cu(II) centre also induce a tilting of the benzene rings
 with respect to the CuO$_2$N$_2$ plane. From the initial angle of
 $\vartheta=34.9$° in CuCCP the tilting due to the H$_2$O ligand is quite
 significant, leading to $\vartheta=42.9$° in Cu(II)-H$_2$O. The NH$_3$
 molecule, by contrast, leads to a lowering of this angle to $\vartheta=31.8$°
 in Cu(II)-NH$_3$.

In order to quantify the effect of the H$_2$O and NH$_3$ ligands on
the electronic properties of CuCCP, we show in
Table~\ref{ligand_hopping} the values of the \mbox{Cu{\textendash}Cu} hopping
integrals calculated with the NMTO downfolding method where the
hopping parameters for the original CuCCP have been included for
comparison. Note that the intrachain \mbox{Cu{\textendash}Cu} coupling is
reduced by a factor of 1.5-3.5 with the inclusion of both ligands.
The reduction is especially significant with NH$_3$. The only
\mbox{Cu{\textendash}Cu} interchain path that is enhanced is $t_1$ which is
between Cu in nearest neighbour chains and has its origin in the
hydrogen bonds between the H of the H$_2$O (NH$_3$) molecule and the O
of the hydroquinone fragments in the chains. Therefore, apart from
these hydrogen bonds, the introduction of ligands isolates the Cu ions
considerably.

\begin{table}
\caption{
Values for the \mbox{Cu{\textendash}Cu} hopping integrals calculated with the
NMTO downfolding method for the relaxed CuCCP, Cu(II)-H$_2$O and
Cu(II)-NH$_3$ structures. The values are given in meV. The subscripts
$i=1,2,3,7,8,12$ denote the $i$th nearest neighbours. See
Fig. \protect\ref{path1}.}

\begin{center}
\begin{tabular}{|c|c|c|c|}\hline
Path & CuCCP & Cu(II)-H$_2$O & Cu(II)-NH$_3$ \\ \hline \hline
$t_1$ & 4 & 8 & 11 \\
$t_2$ & 8 & 7 & 5\\
$t_3$ & 79 & 57 & 22\\
$t_7$ & 5& 1 & 1 \\
$t_8$ & 3 & 0 & 0\\
$t_{12}$ &0 &0 &0\\ \hline
\end{tabular}
\end{center}
\label{ligand_hopping}
\end{table}

\begin{figure}
\begin{center}
\includegraphics[width=0.48\textwidth]{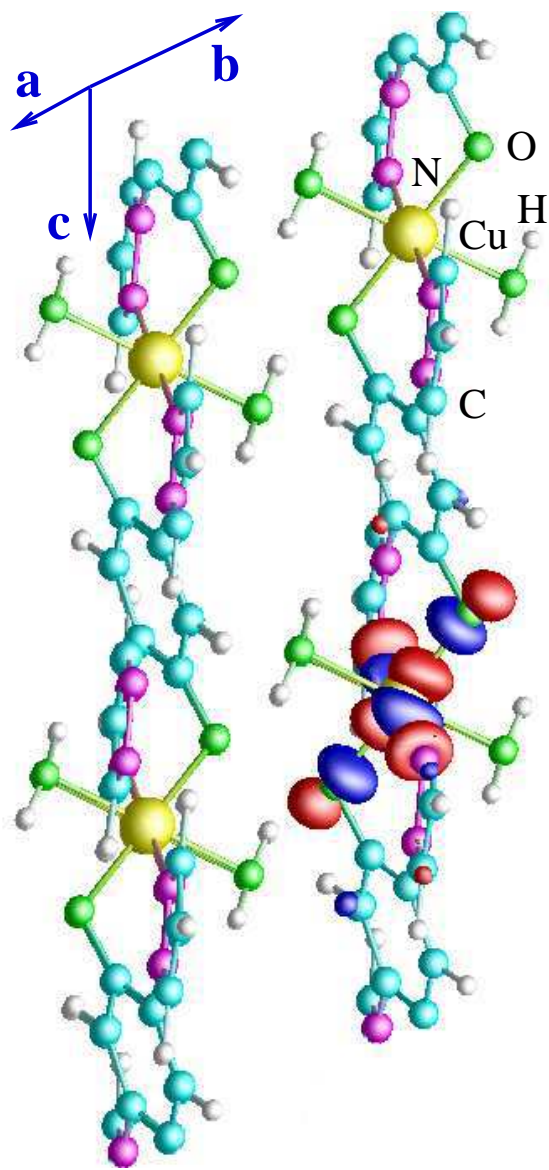}
\end{center}
\caption{
Cu Wannier functions for the Cu(II)-H$_2$O system.}
\label{wannier-ligands1}
\end{figure}

\begin{figure}
\begin{center}
\includegraphics[width=0.6\textwidth]{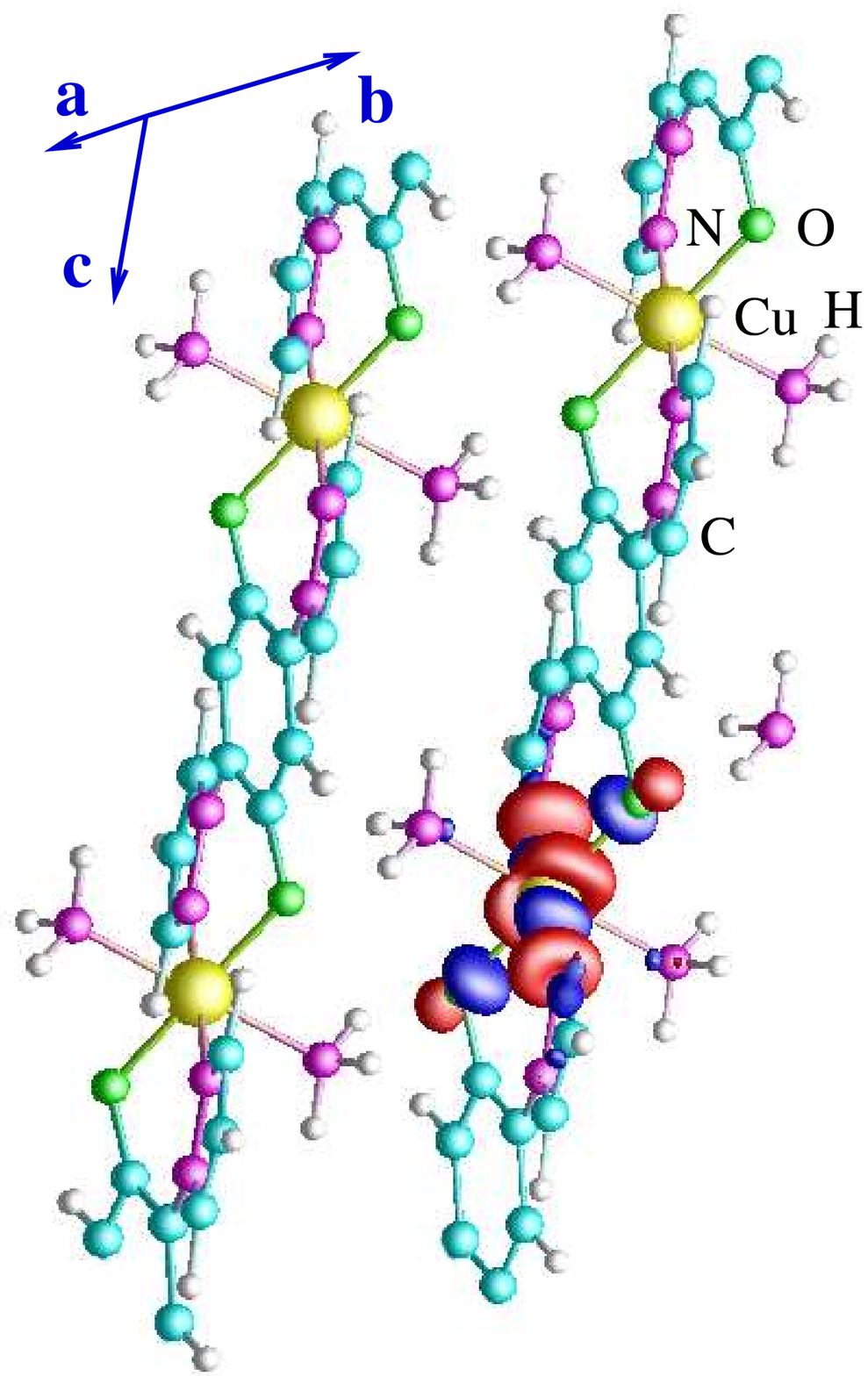}
\end{center}
\caption{
Cu Wannier functions for the Cu(II)-NH$_3$ system.}
\label{wannier-ligands2}
\end{figure}

In the Wannier orbital plot in Figs.~\ref{wannier-ligands1} and
\ref{wannier-ligands2} we can see that the distorted octahedral
environment of the Cu in the Cu(II)-H$_2$O and Cu(II)-NH$_3$
structures induces very little mixing of the Cu~$d_{z²}$ orbital to
the predominant $d_{x²-y²}$. Also note the little contribution of
weight in the hydroquinone ring, in contrast to the previous discussed
systems (see Fig.~\ref{wannier}) which is a manifestation of the
isolated nature of Cu in these structures.  The inclusion of the
H$_2$O and NH$_3$ satellites, however, changes the Cu coordination
only marginally from four in the direction of six, as opposed to our
original motivation for the addition of H$_2$O and NH$_3$ ligands. Cu
remains in the oxidation state of almost $2+$ as observed in our
calculations. While the GGA calculations give a metallic behaviour with
a half-filled predominantly $d_{x²-y²}$ Cu band, inclusion of
correlation effects with LDA+U drive the system to an insulating
state. Therefore the system will remain an insulator.

\section{Summary}\label{sec:summary}

In search of low-dimensional quantum spin systems with tunable
properties, we have proposed and analyzed within an {\it ab initio}
framework and using a combination of different computational methods
various chemical modifications to the Cu-based polymeric coordination
compound CuCCP.  Our goal has been to tune in a controlled way the
magnetic interactions between Cu centres and to test the efficiency
and feasibility of the combination of methods proposed here.  We
pursued two ways of modifying the original CuCCP structure; by
changing the substitution pattern in the linker (hydroquinone) and by
adding ligands to the system.  Following the first scheme we
considered two possible H substitutions in the hydroquinone; an
electron donating group (NH$_2$) and an electron withdrawing group
(CN).  For the second scheme, we considered structures with H$_2$O and
NH$_3$ ligands.  Out of our study we conclude that the NH$_2$
substitution in the hydroquinone is the most effective in order to
enhance the intrachain Cu-Cu interaction in CuCCP while the CN
substitution induces and enhances the interchain interactions in the
system which were either absent or very weak in the original CuCCP
compound.  In contrast, the inclusion of H$_2$O or NH$_3$ ligands has
the effect of isolating the Cu ions.

The effects observed in this study are small, mainly due to the fact
that the coupling constants in these metalorganic materials are weak.
On the other hand, these systems are, due to the smallness of the
coupling constants, of special interest since application of moderate
magnetic fields or pressure can drive the system to a phase
transition.  These systems can be described as Mott-Hubbard insulators
and possibly under application of pressure a metal-insulator
transition could be induced. This will form the basis of our future
work.

Finally, we believe that the combination of methods presented in the
manuscript is efficient for studying the properties of complex systems
per se, and that the computer-designing procedure that we have
employed in the present study provides a plausible route for
manipulating properties related to low-dimensional quantum spin
systems in general.

\newpage

\appendix
\section*{Appendix}
\setcounter{section}{1}

We present here the AIMD relaxed structural data of the various Cu(II)
polymers. Other than CuCCP, the rest are computer designed.

In Table~\ref{crysdata2} we show the fractional atomic positions of the
CuCCP polymer obtained after the relaxation of the system. The
resulting distances between the atoms  after the optimization are
closer to the standard values found in the literature~\cite{allen:87}
but differs from the values for the distances in the experimental
compound~\cite{dinnebier:02}.

\begin{table}
\caption{ 
Fractional atomic coordinates of nonequivalent atoms in the CuCCP
relaxed structure. For the lattice parameters, see Table~\ref{latpam}.}

\begin{center}
\begin{tabular}{|c|c|c|c|} \hline
Atom & x & y & z \\ \hline \hline
Cu & 0.5000 & 0.5000 & 0.5000 \\
O2 & 0.4581 & 0.3495 & 0.6187 \\
C3 & 0.4745 & 0.4243 & 0.7998 \\
C4 & 0.6696 & 0.5869 & 0.9293\\
C5 & 0.0331 & 0.7729 & 0.6797\\
C6 & 0.2206 & 0.8455 & 0.8363 \\
C7 & 0.1024 & 0.7795 & 0.9497\\
C8 & 0.6971 & 0.6615 & 0.1203 \\
N9 & 0.8576 & 0.6739 & 0.8610 \\
N10 & 0.8109 & 0.6692 & 0.6956\\
H11 & 0.1732 & 0.7923 & 0.0815\\
H12 & 0.4193 & 0.9317 & 0.8607\\
H13 & 0.0456 & 0.7875 & 0.5549\\
H14 & 0.8471 & 0.7919 & 0.2121\\ \hline
\end{tabular}
\end{center}
\label{crysdata2}
\end{table}

In Table~\ref{crysdata-CN} we present the relative atomic positions
obtained after relaxation for Cu(II)-CN. The cell parameters were
fixed during relaxation, and the only change observed was the
tilting of the hydroquinone ring as a consequence of the movement
upward of the pyrazolyl rings.

\begin{table}
\caption{
Fractional atomic positions of nonequivalent atoms in Cu(II)-CN
obtained with the PAW method. For the lattice parameters, see
Table~\ref{latpam} (same as CuCCP).}

\begin{center}
\begin{tabular}{|c|c|c|c|} \hline
Atom & x & y & z \\ \hline \hline
Cu & 0.5000 & 0.5000 & 0.5000 \\
O2 & 0.4559 & 0.3460 & 0.6195 \\
C3 & 0.4823 & 0.4252 & 0.7998 \\
C4 & 0.6795 & 0.5881 & 0.9253 \\
C5 & 0.0384 & 0.7593 & 0.6620 \\
C6 & 0.2349 & 0.8314 & 0.8139 \\
C7 & 0.1239 & 0.7694 & 0.9310 \\
C8 & 0.6944 & 0.6601 & 0.1182 \\
C9 & 0.1030 & 0.1811 & 0.7598 \\
N10 & 0.8687 & 0.6739 & 0.8553 \\
N11 & 0.8152 & 0.6676 & 0.6880 \\
N12 & 0.9162 & 0.0575 & 0.6788 \\
H13 & 0.2104 & 0.7793 & 0.0570 \\
H14 & 0.4353 & 0.9135 & 0.8323 \\
H15 & 0.0493 & 0.7681 & 0.5346 \\ \hline
\end{tabular}
\end{center}
\label{crysdata-CN}
\end{table}

In Table~\ref{crysdata-NH3}  we show the same, but for Cu(II)-NH$_3$. 

\begin{table}
\caption{
Fractional atomic positions of nonequivalent atoms in Cu(II)-NH$_3$
obtained with the PAW method. For the lattice parameters, see Table~\ref{latpam}.} 

\begin{center}
\begin{tabular}{|c|c|c|c|} \hline
Atom & x & y & z \\ \hline \hline
Cu & 0.5000 & 0.5000 & 0.5000 \\
O2 & 0.4679 &  0.3798 &0.6285 \\
C3 & 0.4786  & 0.4444 & 0.8064 \\
C4 & 0.6761 & 0.6148 & 0.9103 \\
C5 & 0.0881 & 0.8540 & 0.6283 \\
C6 & 0.2867 &0.9709 &  0.7753 \\
C7 &  0.1520 &  0.8949 &  0.9023 \\
C8 & 0.6932 &  0.6630 & 0.0956 \\
N9 &  0.8861 & 0.7410 &  0.8310 \\
N10 & 0.8489 & 0.7159 & 0.6631 \\
H11 &  0.2228 &  0.9402 & 0.0374 \\
H12 &0.4977  &0.0958 & 0.7892 \\
H13 & 0.0972 & 0.8601 & 0.4997 \\
H14 &0.8313  & 0.7917 & 0.1793 \\
N15 & 0.8587 &  0.4701 & 0.3123 \\ 
H16 & 0.7612 & 0.3507 & 0.2812 \\
H17 &  0.0789 & 0.5407 &  0.3673 \\ 
H18 & 0.9088  & 0.4872 & 0.1907 \\  \hline
\end{tabular}
\end{center}
\label{crysdata-NH3}
\end{table}

The structural data for the Cu(II)-NH$_2$ and Cu(II)-H$_2$O systems
were already presented in Ref.~\cite{jeschke:06}.

\ack

We thank M.~Wagner for useful discussions and T.~Kretz for providing
the information about the aminophenol and o-cyanophenol. This work was
financially supported by the Deutsche Forschungsgemeinschaft (DFG)
through the Forschergruppe FOR 412 and H.O.J. gratefully acknowledges
support from the DFG through the Emmy Noether Program. We gratefully
acknowledge support by the Frankfurt Center for Scientific Computing.
T.S.D and B.R acknowledge MPG-India partnergroup program for the
collaboration.


\begin{thebibliography}{51}
\bibitem{note} 
We have performed non-spin polarized DFT calculations - in contrast to
spin-polarized ones - for the structure relaxation since the energy
associated with magnetism is much smaller than the cohesion energy and
therefore, possible changes on the atomic positions due to the
magnetic energy can be assumed to be negligible.
\bibitem{convergence} 
The calculations were checked for convergence within the two choices
of basis sets (FPLAPW, NMTO).
\bibitem{comment_U} 
Note that this is a very rough U$_\mathrm{eff}$ estimate since we have
here Cu surrounded by two O and two N atoms and the
\mbox{Cu{\textendash}Cu} path is via a complicated organic
linker. Typical values of $U_\mathrm{eff}$ for Cu oxides is 4-6 eV.
\end{thebibliography}
\end{document}